\documentclass[prd,showkeys,floatfix,nofootinbib,
               preprint,12pt,
               tightenlines,fleqn]{revtex4-1} 

\usepackage{amsmath,amsthm,amssymb,revsymb,graphicx,dcolumn}
\usepackage{leftidx}

\usepackage{marginnote}
\usepackage{enumitem}

\newcommand{\beq}{\begin{equation}}
\newcommand{\eeq}{\end{equation}}
\newcommand{\beqa}{\begin{eqnarray}}
\newcommand{\eeqa}{\end{eqnarray}}
\newcommand{\bsubeqs}{\begin{subequations}}
\newcommand{\esubeqs}{\end{subequations}}

\usepackage{extarrows,tikz,tikzscale,adjustbox}
\usetikzlibrary{matrix}

\usepackage{etoolbox}
\makeatletter
\patchcmd{\frontmatter@RRAP@format}{(}{}{}{}
\patchcmd{\frontmatter@RRAP@format}{)}{}{}{}
\renewcommand\Dated@name{}
\makeatother

\begin{document}

\title[]
      {Coordinate- and spacetime-independent quantum physics}      
\author{V.A.\;Emelyanov}
\email{viacheslav.emelianov@rwth-aachen.de}
\author{D.\;Robertz}
\email{daniel.robertz@rwth-aachen.de}
\affiliation{Lehrstuhl\;f\"ur\;Algebra\;und\;Zahlentheorie\\
RWTH\;Aachen\;University\\
Pontdriesch\;14/16\\
D-52062\;Aachen\\
Germany}

\begin{abstract}
\vspace*{2.5mm}\noindent
The concept of a particle is ambiguous in quantum field theory.~It is generally agreed that
particles depend not only on spacetime,~but also on coordinates used to parametrise spacetime
points.~One of us has in contrast proposed a coordinate-frame-independent model of quantum particles~within~the
framework of quantum field theory in curved spacetime.~The aim of this article is to present a scalar-
field-equation
solution that is not only a zero-rank tensor under general coordinate transformations, 
but also common for anti-de-Sitter, de-Sitter, closed and open Einstein static universes.~Moreover,~it
locally reduces to a Minkowski plane-wave solution and is non-perturbative in curvature.\,The former 
property makes it suitable for the standard applications of quantum theory in particle physics,~while the latter allows then to~gain~insights~into quantum physics in the strong-gravity regime.
\end{abstract}

\maketitle

\section{Introduction}

In the lack of experimental data on the metric tensor describing the universe geometry,~it
is common to approximate \emph{local} patches of the universe geometry by spacetimes admitting
symmetry groups.~Such an approximation~depends on a length scale:~The~de-Sitter~geometry
best fits to observations at cosmological scales, while the Kerr geometry is appropriate~for~the
description of the Earth's gravitational field.~Moreover,~at small-enough scales and away from singularities,~all of them
reduce to Minkowski spacetime.~It is due to the Einstein equivalence
principle that makes general relativity locally compatible with special relativity~\cite{Casola&etal}.

In quantum field theory,~spacetime symmetries are used by selecting local-field operators
to create particles out of quantum vacuum~\cite{Birrell&Davies}.~It successfully works in~the case of Minkowski
spacetime with isometries generated by elements of the Poincar\'{e} algebra.~Namely,~quantum
particles are linked to irreducible unitary representations~of~the Poincar\'{e} group,~according
to the Wigner classification~\cite{Weinberg}.~However, the universe geometry cannot be globally~modelled~by
Minkowski spacetime.~In other words,~the Wigner classification only applies locally.~This~is~in
agreement with collider-physics experiments.~There~are,~though,~no~experimental~data~which would tell
us~that,~for instance, an analogous classification based on isometries
of~Kerr~spacetime must replace the Wigner classification in the Earth's gravitational field.~Nevertheless,~it
is generally agreed that the concept of a quantum particle depends not only on spacetime,~but also on
coordinates used to parametrise spacetime points~\cite{Birrell&Davies,Schroedinger,Parker,Fulling,Hawking,Davies,Unruh,Gibbons&Hawking}.

Still,~a superposition of plane waves in a local inertial frame properly describes quantum
particles in the Earth's gravitational field.~This follows from the~Colella-Overhauser-Werner
experiment~\cite{Colella&Overhauser,Colella&Overhauser&Werner}.~The subsequent outcome of
the Bonse-Wroblewski~experiment~\cite{Bonse&Wroblewski}~shows that
interference of quantum particles induced by homogeneous gravity
is indistinguishable from that induced by uniform acceleration,~and vice versa~\cite{Nauenberg}.~These
observations~lead to~the idea that quantum particles must be modelled by wave functions
which are,~first,~locally~given by plane-wave superpositions.~Second,~wave functions must be~tensors~with 
respect to general coordinate
transformations.~This is logically~in accord~with~the Einstein field equations which require that
(quantum)
matter curve spacetime via energy-momentum
\emph{tensor}~\cite{Emelyanov-2020,Emelyanov-2021,Emelyanov-2022a,Emelyanov-2022b}.

The observable-universe geometry does not only vary over length scales,~but also from~one
local region to another, each of which could be approximated by spacetimes with a non-trivial 
group of isometries.~Still,~(quantum) particles coming to the Earth from distant~regions~of~the
Universe are identified with those from the Standard Model of elementary particle physics~\cite{Weinberg}.
It thus follows that a particular spacetime isometry approximately realised in a certain region of
the Universe should be irrelevant for the definition of a quantum-particle concept.~Indeed,
in the semi-classical limit, one should be able to reproduce classical-physics results by which
the classical-particle concept is oblivious to the geometry of a given spacetime.

The article's aim~is~to~present~a covariant scalar-field-equation solution being common
for a set of spacetimes.~Specifically, we consider non-perturbatively five spacetimes at once:~Anti-
de-Sitter (AdS), de-Sitter (dS), Minkowski, closed and open Einstein static universes~(ESUs).
The main purpose is to demonstrate that there exists a single quantum-particle~notion~for~all 
these spacetimes,~even though their global isometry groups differ from each other.

Throughout, we use natural units $c = G = \hbar = 1$, unless otherwise stated.

\section{Physical motivation}
\label{sec:pm}

\subsection{Field quantisation in curved spacetime}

Quantum field theory is the basic formalism used to model phenomena involving particles.
In contrast to quantum mechanics,~this formalism deals with distribution-valued operators,
known as field operators.~These give rise to the concept of a field operator algebra.~In~physics,
it is necessary to choose a particular Hilbert-space representation of such an algebra,~which
gives rise to the concept of a quantum particle.~However,~the Stone-von Neumann
theorem~is invalid in quantum field theory as a system with uncountably many degrees of freedom.~This
implies that different Hilbert-space representations
may be~unitarily~inequivalent~\cite{Haag}.~This aspect of the formalism finds its application by
the description~of phase transitions~\cite{Umezawa}.

Quantum field theory surpasses quantum mechanics in the sense that the former includes
the principles of special relativity.~The concept of a quantum particle
should accordingly~be independent of inertial frames of reference.~This is achieved via
the identification of particles' states with irreducible unitary representations of the Poincar\'{e}
group.~Therefore,~the~isometry group of Minkowski spacetime -- the Poincar\'{e} group --
determines the unique Hilbert-space representation.~This choice agrees with collider-physics
experiments~\cite{Weinberg}.

In the framework of general relativity, the observable Universe is a non-Minkowski space-
time.~It is generally agreed that the isometry group of a given curved spacetime distinguishes
a Hilbert space which has applications in particle physics.~Still,~even the
isometry groups~of the maximally symmetric curved spacetimes~--~AdS and dS~--~are not enough to pick
a~unique Hilbert space.~It is then also generally agreed that the concept of a quantum particle
depends on the concept of observer's time.~The relativity of time leads in turn to the ambiguity of the concept~of
a quantum particle~in quantum field theory in
curved spacetime.~This hypothesis led~to 
predictions~\cite{Schroedinger,Parker,Fulling,Hawking,Davies,Unruh,Gibbons&Hawking} which,~though,~lack experimental
confirmations for the moment.

However,~there are experimental data on quantum effects due to the Earth's gravitational
field.~For example,~a non-trivial interference pattern~observed~via
overlapping of two~beams~of thermal neutrons,~moving at different altitudes with respect to the Earth's
surface,~forms~due to the free-fall
acceleration $g_\oplus$~\cite{Colella&Overhauser,Colella&Overhauser&Werner}.~This experimental result
can be generalised in theory to any weak gravitational field~\cite{Stodolsky}.~This implies that
the interference pattern is owing~to~the difference in proper time accumulated by the neutrons
traveling along different paths.~This generalisation also agrees with the gravitational
Aharonov-Bohm effect~\cite{Audretsch&Laemmerzahl},~which has been recently 
observed~in~\cite{Overstreet&etal}.

Particles should accordingly be related to their proper time,~rather than to observer's~time.
It~manifests itself in Minkowski spacetime through time dilation -- a mean lifetime of unstable 
particles should increase in a laboratory by increasing their relative velocity to that.~In~fact,
this agrees with the observation
of~cosmic-ray muons at the Earth's surface~\cite{Rossi&Hall},~whereas~the 
Earth's gravitational field locally plays no role,~as in collider-physics experiments~\cite{Weinberg}.~Taking
these empirical results as a clue to modelling quantum particles in curved spacetime,~we~have proposed
that particles' states in curved spacetime should \emph{locally} match those based on the 
representations of the Poincar\'{e} group~\cite{Emelyanov-2020,Emelyanov-2021,Emelyanov-2022a,Emelyanov-2022b}.

\subsection{Einstein's equivalence principle}

The Standard Model of particle physics uses the formalism of quantum field theory~for~the
description of scattering processes and decay rates in collider physics.~The formalism~assumes
no gravitational field -- the observable Universe is approximated by Minkowski spacetime~\cite{Weinberg}.
However,~the~Einstein
equivalence principle allows only to consider local Minkowski frames~in the Universe, meaning
the metric tensor can \emph{only locally} be brought to the form
\beqa
g_{ab}(y)\big|_{\textrm{Universe}} &=& \eta_{ab} - \frac{1}{3}\,R_{acbd}(0)\,y^cy^d +
\textrm{O}(y^3)\,,
\eeqa
where $y$ are Riemann normal coordinates introduced at some non-singular point ($y^a = 0$) in the Universe
\cite{Petrov},~and $R_{abcd}(0)$ is the Riemann curvature tensor computed at that point.~The 
coordinate-dependent corrections to the Minkowski metric tensor~$\eta_{ab}$
can~be~neglected~whenever deviations from $y^a = 0$ are much less than a local curvature~length
at $y^a = 0$.

The formalism of quantum field theory and its applications in collider physics
successfully work
in local Minkowski frames introduced at the Earth's surface.~This is an empirical~fact.~In
local Minkowski frames, scattering processes and decay rates are computed by making use~of
the Lehmann-Symanzik-Zimmermann reduction formula, connecting $S$-matrix elements~with
time-ordered Green's functions~\cite{LSZ}.~The reduction formula basically relies~on~the~concept~of
asymptotic states, which enter the $S$-matrix and correspond to particles moving~at~constant
momenta.~Such states for a Klein-Gordon field, $\hat{\Phi}(y)$, are defined through creation operators
by use of the Klein-Gordon inner product~\cite{Weinberg,LSZ,Srednicki}:
\beqa\label{eq:adagger-minkowski}
\hat{a}^\dagger(K,\Sigma) &\approx& 
-i{\int_{\Sigma}}d\Sigma(y)\,n^a\Big({\exp_K(y)}\,\partial_a\hat{\Phi}^\dagger(y)
-\hat{\Phi}^\dagger(y)\,\partial_a\exp_K(y)\Big)\,,
\eeqa
where the approximation sign is supposed to emphasise that 
the observable Universe~is~a~non-
Minkowski spacetime,
$n^a$ is a future-directed unit four-vector orthogonal
to Cauchy's surface $\Sigma$~and $d\Sigma(y)$
denotes the volume element in $\Sigma$,~and
\beqa\label{eq:pws}
\exp_K(y) &\equiv& {\exp}{\big({-}i\eta_{ab}K^a y^b\big)}\,,
\eeqa
where $K$ is a on-mass-shell four-momentum,~namely~$\eta_{ab}K^aK^b = M^2$.~The definition~\eqref{eq:adagger-minkowski}~gives
\beqa\label{eq:adagger-minkowski-e}
\hat{a}^\dagger(K,\Sigma_f) - \hat{a}^\dagger(K,\Sigma_i) &\approx& -i{\int}d^4 y
\exp_K(y) \big(\eta^{ab}\partial_a\partial_b + M^2\big) \hat{\Phi}^\dagger(y)\,,
\eeqa
where the integration is over the space-time volume with the space-like boundaries $\Sigma_f$ and~$\Sigma_i$.
Strictly speaking,~this formula needs the consideration of a wave packet being a superposition of
$\exp_K(y)$,~see Sec.~5 in~\cite{Srednicki} for further details.~If
$\big(\eta^{ab}\partial_a\partial_b + M^2\big) \hat{\Phi}(y) = 0$ holds, then the
operator $\hat{a}^\dagger(K,\Sigma)$ is independent of the Cauchy surface $\Sigma$.~If otherwise,
which is the case~in interacting~(non-linear)~quantum field theory,~$\hat{a}^\dagger(K,\Sigma)$ changes with
time.~The~formula~\eqref{eq:adagger-minkowski-e}~is used to express 
the $S$-matrix in terms of time-ordered Green's functions,~e.g. see~\cite{Srednicki}.
 
Asymptotic states model particles being far away from each
other,~such that their mutual interaction~is negligible.\,Such states are independent of $\Sigma$~till
particles~get~closer~to~each~other. Accordingly,~an asymptotic state reads
$|K\rangle = \hat{a}^\dagger(K)|\Omega\rangle$~with~$|\Omega\rangle$
being quantum vacuum,~i.e. $\hat{a}(K)|\Omega\rangle = 0$.~This state describes a particle
moving along $y^a = (K^a/M)\,\tau$,~where $M$ denotes the particle's mass and
$\tau$ is proper time.\,The plane wave\,$\exp_K(y)$\,turns into~$\exp({-}iM\tau)$~on~the
particle's trajectory,~meaning the particle's phase is proportional to its proper time.~The~same 
trajectory~looks differently if considered relative to the Earth's surface.~However,~proper~time
is invariant under general coordinate transformations,~i.e.~independent of observer's frame~of
reference,~but varies depending on a particle's trajectory.~In terms of observer's time,~being 
at rest with respect to the Earth's surface, particles
moving at different altitudes experience different gravitational time 
dilations:~$\tau(h) \approx (1 + g_\oplus h)\,\tau(0)$,~where $h$ is the height relative~to the Earth's surface.~This
general-relativity effect combined with
particles' phase~proportional to proper time
explains the quantum interference induced by gravity observed in~\cite{Colella&Overhauser&Werner}.

The plane-wave solution $\exp_K(y)$ is,\,however,\,an approximate
solution~of~the~(linear)~Klein-Gordon equation in the Universe.~The basic idea in quantum field
theory in curved spacetime consists in the choice of a substitution for $\exp_K(y)$ based on the concept
of observer's time~\cite{Birrell&Davies}.
However,~the
definition~\eqref{eq:adagger-minkowski} admits~a generalisation
by use of an exact solution $\textrm{sol}_K(y)$:
\beqa\label{eq:adagger-universe}
\hat{a}^\dagger(K,\Sigma) &\equiv&
-i{\int_{\Sigma}}d\Sigma(y)\,n^a\Big({\textrm{sol}_K(y)}\,\nabla_a\hat{\Phi}^\dagger(y)
-\hat{\Phi}^\dagger(y)\,\nabla_a\textrm{sol}_K(y)\Big)\,,
\eeqa
where $\partial_a$ has been replaced by the covariant derivative $\nabla_a$, such that
\beqa
\textrm{sol}_K(y) &\xrightarrow[C \,\to\, 0]{}& \exp_K(y)\,,
\eeqa
where ``$C \to 0$" assumes that space-time ($C$) curvature is neglected.~This~holds,~in~practice, 
whenever a wave packet built out of a superposition of $\textrm{sol}_K(y)$ has a spatial size being
much smaller than the local curvature length (which is roughly $10^{11}\,\textrm{m}$ at the Earth's
surface).~Thus,
the generalisation from local Minkowski frames to the observable Universe
reduces~in~the~end to determining $\textrm{sol}_K(y)$ in the Universe which is locally approximated
by $\exp_K(y)$.

\subsection{General covariance}

The concept of proper time
is invariant under general coordinate transformations.~Besides,
particles carry energy and momentum,~originating from particles' energy-momentum~tensor.
The~latter in turn is part of Einstein's field equations.\,It then logically follows that particles
must be modelled by wave packets transforming as tensors~under 
coordinate~transformations.
It particularly assumes that $\textrm{sol}_K(y)$ must be a scalar.~Specifically,
for a particle created~by~\eqref{eq:adagger-universe}
out of quantum vacuum,~$|\Omega\rangle$,~to possess an energy-momentum
tensor,~rather~than~an energy-momentum
matrix-valued function, the operator $\hat{a}^\dagger(K,\Sigma)$ must be a zero-rank
tensor which requires $\textrm{sol}_K(y)$ to be a scalar.

In classical theory,~particles' states are characterised by initial position and momentum:~$X$
and $P$.~In quantum theory,~particles' states are characterised by wave~packets.~The~latter~may
be reduced to the former description if wave packets also carry information about initial position 
and momentum.~While $P$ may enter a wave packet via a superposition~of~$\textrm{sol}_K(y)$~with
a weight depending on $\eta_{ab}K^aP^b$ and having its peak at $K = P$,~$X$ must~be~tensorially~coupled
with the wave-packet argument $x$.~This can be achieved through geodesic distance
\beqa
\sigma(x,X) &\equiv& \frac{1}{2}\,{\int\limits_{X \,=\, x(0)}^{x \,=\, x(1)}} g_{\mu\nu}(x(s))\, \dot{x}^\mu(s)\,\dot{x}^\nu(s)\,ds\,,
\eeqa
where dot denotes the differentiation with respect to $s$, which gives the one-half the square of the distance
along the geodesic between $x$ and $X$~\cite{DeWitt}.~Wave packets may in the end~depend 
on their
argument $x$ through $\sigma(x,X)$ and its covariant derivatives only~\cite{Emelyanov-2020}.

In the Riemann frame, $2\sigma(x,X)$ equals $\eta_{ab}y^ay^b$, where $X^a$ corresponds to
$y^a =0$~\cite{Ruse}.~Thus, 
$\textrm{sol}_K(y)$ is a scalar depending on $y^a$, $K^a$, $\eta_{ab}$ and
the curvature tensors at $y^a = 0$.

\subsection{Mathematical and physical reasoning for AdS, dS, closed and open ESUs}

In general,~there are infinitely many covariant variables which can be constructed~by~use~of
$y^a$, $K^a$, $\eta_{ab}$ and their contractions with the curvature tensors.~This makes it hardly
feasible~in general to obtain $\textrm{sol}_K(y)$ which is non-perturbative in curvature.

However,~in the case of AdS and dS spacetimes, we have
\beqa\label{eq:riemann-(a)ds}
R_{abcd}(0)\big|_\textrm{(A)dS} &\propto& \eta_{ac}\eta_{bd} - \eta_{ad}\eta_{bc}\,,
\eeqa
where the coefficient of proportionality depends on which of the spacetimes we consider.~This
means that $\eta_{ab}K^ay^b$ and $\eta_{ab} y^a y^b$ exhaust all independent covariant
variables on which~$\textrm{sol}_K(y)$ may depend~\cite{Emelyanov-2020}.~In contrast,~closed~and~open ESUs are characterised by
\beqa\label{eq:riemann-esus}
R_{abcd}(0)\big|_\textrm{ESUs} &\propto& 
(\delta_{ac} - \eta_{ac})(\delta_{bd} - \eta_{bd}) - (\delta_{ad} - \eta_{ad})(\delta_{bc} - \eta_{bc})\,,
\eeqa
where $\delta_{ab}$ is the Kronecker delta and the proportionality coefficient depends on the universes.
It will turn out that there exist only three independent covariant
variables in the ESUs.~These circumstances are the basic mathematical reason why we below
consider these spacetimes.\;\;

From a physics point of view,~the observable Universe is modelled by de-Sitter~spacetime at cosmological scales,~according to the~Standard Model of  Cosmology~\cite{Mukhanov}.~It~is~also~applied
in combination with~quantum field theory~to account for 
the origin of inhomogeneities~in~the Universe formed over~cosmic inflation~\cite{Mukhanov&Chibisov}.~Anti-de-Sitter
spacetime is part~of~the~conjectured relationship between quantum gravity\,as\,modelled by\,string 
theory~and~conformal~field~theory
in Minkowski spacetime~\cite{Maldacena,Witten,Gubser&Klebanov&Polyakov}.~Although the Einstein universes
have no applications~in~the Standard Model of Cosmology, a closed~Einstein universe might be considered~as~a 
static~limit of the Oppenheimer-Snyder model of gravitational collapse~\cite{Oppenheimer&Snyder}.~Furthermore,
closed Einstein static spacetime is given by a
$d$-dimensional sphere,~where at each~of~its~points~there~is~time evolution.~Thus,~a $3$-dimensional
closed ESU may be mimicked in a table-top experiment~held
at~the International
Space Station,~by trapping particles to a $2$-dimensional sphere.~We~shall elaborate on this
setup below.~Finally,~open ESU can be mapped onto AdS,~as will be shown shortly, which
might be of interest in light of~\cite{Maldacena,Witten,Gubser&Klebanov&Polyakov}. 

\section{Metric tensor in Riemann normal coordinates}
\label{sec:metric-tensor}

$\textrm{dS}_d$ is a hyperboloid embedded in $(d+1)$-dimensional
Minkowski~spacetime~\cite{Birrell&Davies}.~By~making use of this fact and the fact that great circles
correspond to geodesics 
on a sphere,~we~find\;\;
\bsubeqs\label{eq:ds-esu-gds}
\beqa
2\sigma(x,X)\big|_{\textrm{dS}_d} &=& \frac{1}{H^2}\,{\textrm{arccosh}^2}{\left(1 + 
\frac{\frac{1}{2}H^2(x-X)^2}{\left(1 -\frac{1}{4}H^2x^2\right)\left(1 -\frac{1}{4}H^2X^2\right)}\right)},
\\[1mm]
2\sigma(x,X)\big|_{\textrm{CESU}_d} &=& (x^0-X^0)^2 - 
a^2{\textrm{arccos}^2}{\left(1 - 
\frac{\frac{1}{2}\frac{1}{a^2}(\boldsymbol{x}-\boldsymbol{X})^2}{\left(1 + \frac{1}{4}\frac{1}{a^2}\boldsymbol{x}^2\right)\left(1 + \frac{1}{4}\frac{1}{a^2}\boldsymbol{X}^2\right)}\right)},
\eeqa
\esubeqs
where $H$ and $a$ denote,~respectively,~a Hubble parameter and a radius of the spatial section of a closed
ESU,~and,~by definition, $x^2 \equiv \eta_{\mu\nu}x^\mu x^\nu$ and $\boldsymbol{x}^2  \equiv \delta_{ij}x^i x^j$,~where
Greek-letter~indices run from $0$ to $d-1$ and $i,j \in \{1,\dots,d-1\}$.~These geodesic distances
allow us to express $x$
through Riemann normal coordinates $y$ as follows~\cite{Ruse}:
\bsubeqs
\beqa\label{eq:rnc-gd}
y^a(x) &=& - e_\mu^a(X)g^{\mu\nu}(X)\,\frac{\partial}{\partial X^\nu}\,\sigma(x,X)\,,
\eeqa
where $e_{\mu}^a(x)$ is a $d$-bein field, satisfying
\beqa
\eta_{ab}\,e_\mu^a(x)e_\nu^b(x) &=& g_{\mu\nu}(x)\,,
\eeqa
and $g_{\mu\nu}(x)$ follows from $\sigma(x,X)$:
\beqa
g_{\mu\nu}(x) &=& - \lim_{X\,\to\,x}\frac{\partial}{\partial x^\mu} \frac{\partial}{\partial X^\nu}\,\sigma(x,X)\,.
\eeqa
\esubeqs

\begin{figure}
\adjustbox{width=16.0cm}{\centering
\begin{tikzpicture}
  \matrix (m) [matrix of math nodes,row sep=3em,column sep=4em,minimum width=2em]
  {
     \text{CESU}_{d + 1} & \text{OESU}_{d + 1} \\
     \text{dS}_{d} & \text{AdS}_{d} \\};
  \path[-stealth]
    (m-1-1) edge [->] node [left] {$\text{\scriptsize $\mathcal{DR}\circ\mathcal{AC}$}$} (m-2-1)
            edge [<->] node [above] {$\text{\scriptsize $\mathcal{AC}$}$} (m-1-2)
    (m-2-1.east|-m-2-2) edge [<->] node [below] {$\text{\scriptsize $\mathcal{AC}$}$}
            node [above] {} (m-2-2)
    (m-1-2) edge [->] node [right] {$\text{\scriptsize $\mathcal{DR}\circ\mathcal{AC}$}$} (m-2-2);
\end{tikzpicture}
\begin{tikzpicture}[baseline=-1.335cm]
  \matrix (m) [matrix of math nodes,row sep=3em,column sep=4em,minimum width=2em,scale=2]
  {
     {\textrm{R}}{\times}{\textrm{SO}(\scalebox{0.9}{$d\,{+}\,1$})} & {\textrm{R}}{\times}{\textrm{SO}(\scalebox{0.9}{$1,d$})} \\
     \textrm{SO}(\scalebox{0.9}{$1,d$}) & \textrm{SO}(\scalebox{0.9}{$2,d\,{-}\,1$})  \\};
  \path[-stealth]
    (m-1-1) edge [->] node [left] {$\text{\scriptsize $\mathcal{DR}\circ\mathcal{AC}$}$} (m-2-1)
            edge [<->] node [above] {$\text{\scriptsize $\mathcal{AC}$}$} (m-1-2)
    (m-2-1.east|-m-2-2) edge [<->] node [below] {$\text{\scriptsize $\mathcal{AC}$}$}
            node [above] {} (m-2-2)
    (m-1-2) edge [->] node [right] {$\text{\scriptsize $\mathcal{DR}\circ\mathcal{AC}$}$} (m-2-2);
\end{tikzpicture}}
\caption{Left:\,A\,closed\,(C)\,Einstein static universe\,(CESU)\,can be mapped onto an open\,(O)\,Einstein static
universe (OESU) through analytic continuation ($\mathcal{AC}$) of $\sqrt{-R}$ to $i\sqrt{-R}$, where $R$ is the
Ricci scalar.~The map is invertible and works for the pair of a dS and  an AdS as well.~Besides,~a CESU can also be mapped into a de-Sitter spacetime
through dimensional reduction and analytic continuation ($\mathcal{DR}\circ\mathcal{AC}$).~This is accomplished
by reducing the CESU to its spatial section and by promoting one of its spatial (Riemann normal)
coordinates to an imaginary
variable.~Its imaginary part turns~into a time variable
in the dS.~This
procedure also works for the map from an OESU to an AdS.~Right:
The commutative diagram is represented through
the maps of isometry groups of the corresponding spacetimes.}\label{fig:1}
\end{figure}

\noindent From $g_{\mu\nu}(x) \to g_{ab}(y)$ under the coordinate 
transformation
$x^\mu \to y^a$ for dS and ESU,~we~find
for all the spacetimes under consideration that
\beqa
g_{ab}(y) &=& \eta_{ab} - \frac{{\sinh}^2{\sqrt{-R_{ab}\,y^a y^b/(1+2\alpha)}}-(-R_{ab}\,y^a y^b/(1+2\alpha))}
{\left(-R_{ab}\,y^a y^b/(1+2\alpha)\right)^2}\,
R_{acbd}\,y^c y^d\,,
\eeqa
where $R_{abcd}$ and $R_{ab}$ are,~respectively,~Riemann and Ricci tensors 
at $y^a = 0$,~and
\beq\label{eq:alpha}
\alpha \,\equiv\,
\left\{
\begin{array}{ll}
(d - 2)/2  & \textrm{for AdS and dS universes}\,, \\[2mm]
(d - 3)/2 & \textrm{for closed and open ESUs}\,,
\end{array}
\right.
\eeq
which may be expressed via the ratio of $R_{ab}R^{ab}$ and Kretschmann's scalar $R_{abcd}R^{abcd}$.

Comparing the geodesic distances~\eqref{eq:ds-esu-gds},~we observe
\beqa
\sigma(x,X)\big|_{\textrm{CESU}_{d+1}} &\to& \sigma(x,X)\big|_{\textrm{dS}_d}
\eeqa
assuming $x^0 = X^0$, $x^j \to i x^0$ and $X^j \to i X^0$ for one of
$j \in \{1,\dots,d\}$~in~a~$\textrm{CESU}_{d+1}$,~while~$1/a$ turns into a Hubble parameter
in $\textrm{dS}_d$.~This map is a composition of~analytic continuation~and
dimensional reduction.~This map can also be observed at the level of the metric tensor~$g_{ab}(y)$.
Furthermore,~this can be extended to include $\textrm{AdS}_d$ and $\textrm{OESU}_{d+1}$,~see fig.~\ref{fig:1} for details.~These observations play a key role for constructing a single non-perturbative $\textrm{sol}_K(y)$ below.

\section{Klein-Gordon equation in covariant variables}

For the sake of simplicity,~we consider a scalar field which satisfies
a massive Klein-Gordon equation with conformal coupling to gravity~\cite{Birrell&Davies}:
\beqa\label{eq:ofeq}
\left(g^{ab}(y)\nabla_a\nabla_b + M^2 - \frac{d-2}{4(d-1)}\,R(y)
\right)\textrm{sol}_K^{(\alpha)}(y) &=& 0\,,
\eeqa
where $\nabla_a$ is the covariant derivative, $M > 0$ stands for the mass parameter,~and $R(y)$ is the Ricci scalar at the point $y$.~We
intend to find a solution of this equation,~which is a zero-rank tensor with respect to general
coordinate transformations~\cite{Emelyanov-2020,Emelyanov-2021,Emelyanov-2022a,Emelyanov-2022b}.

It follows from the isometry groups of the spacetimes under consideration (see fig.~\ref{fig:1},~right),
or directly from~\eqref{eq:riemann-(a)ds} and~\eqref{eq:riemann-esus},
that
there exist only four basic covariant space-time variables:\;\;\;\;
\bsubeqs
\beqa
v_1(y) &\equiv& +\eta_{ab}\,K^a y^b\,,
\\[2mm]
v_2(y) &\equiv& +2 (1+\alpha)R_{ab}\,K^a y^b/R\,,
\\[2mm]
v_3(y) &\equiv& - R_{ab}\,y^a y^b/(1+2\alpha)\,,
\\[2mm]
v_4(y) &\equiv& - R\,\eta_{ab}\,y^a y^b/2(1+\alpha)(1+2\alpha)\,,
\eeqa
\esubeqs
where $K = K^a\partial_a$ is a momentum $d$-vector defined at $y^a = 0$.~This momentum vector~needs~to be introduced~as we wish to find a Klein-Gordon-equation solution which locally behaves~as~a plane wave,
in accord with the momentum-space representation in quantum theory.~In~terms of these variables,~we
then find from
\beqa\label{eq:c-phi}
\textrm{sol}_K^{(\alpha)}(y) &\equiv&  \frac{1}{4\pi}\,\frac{{\sinh^\alpha}{\eta(v)}}{{\sinh^\alpha}{\zeta(v)}}\,
\frac{\phi^{(\alpha)}(v)}{\left(\scalebox{0.9}{$i\sqrt{\gamma(1-\gamma)}$}\right)^{\alpha}}\,
\eeqa
that the scalar-field equation~\eqref{eq:ofeq} turns by use of elimination theory into
\beqa\label{eq:cfeq}
\left(
\partial_\eta^2
+ \frac{\gamma(1-\gamma) + 
(\nu^2 - \mu^2)\big(\partial_\chi^2 + \xi \partial_\xi^2 +1\big)}{{\sinh^2}{\eta}}
- \partial_\zeta^2 - \frac{\alpha(1-\alpha)}{{\sinh^2}{\zeta}}
\right) \phi^{(\alpha)}(v) &=& 0\,,
\eeqa
where we have defined the following variables:
\bsubeqs\label{eq:mcv}
\beqa
\eta(v) &\equiv& \ln \tanh \frac{\sqrt{v_3}}{2}\,,
\\[0mm]
\zeta(v) &\equiv& 
\ln \frac{\sqrt{v_2^2-\mu^2 v_3}+v_2}{\sqrt{\mu^2v_3}}\,,
\\[1mm]
\chi(v) &\equiv& - v_1 + v_2\,,
\\[2mm]
\xi(v) &\equiv& \frac{1}{4}\left(\big(\nu^2-\mu^2\big)\big(v_4 - v_3\big)-
\big(v_1 - v_2\big)^2\right),
\eeqa
\esubeqs
where $\eta(v)$ and $\zeta(v)$ generalise (34) in~\cite{Emelyanov-2020} to the non-de-Sitter 
spacetimes,~and
\bsubeqs
\beqa
\nu^2 &\equiv& -2(1+\alpha)(1+2\alpha)\,\frac{M^2}{R}\,,
\\[1mm]
\mu^2 &\equiv& -4(1+\alpha)^2(1+2\alpha)\,\frac{R_{ab}K^a K^b}{R^2}\,,
\\[1mm]
\gamma &\equiv& \frac{1}{2}\left(1-i\sqrt{4\mu^2 - \frac{(1+2\alpha)(2\alpha-d+3)}{(d-1)}}\right).
\eeqa
\esubeqs

The variables~\eqref{eq:mcv} can be computed in the spacetimes we
study.~In particular,~we find~by
use of~\eqref{eq:riemann-(a)ds} and~\eqref{eq:riemann-esus} that
\bsubeqs\label{eq:van-var}
\beqa
\chi|_\textrm{(A)dS} &=& \xi|_\textrm{(A)dS} \;=\; 0\,,
\\[2mm]
\xi|_\textrm{ESUs} &=& 0\,.
\eeqa
\esubeqs
There are,~thereby,~two independent covariant variables in
(A)dS~\cite{Emelyanov-2020}~and~three~in ESUs.~This
result can be understood as follows.~The Riemann tensors in~\eqref{eq:riemann-(a)ds}
and~\eqref{eq:riemann-esus}~can~be~expressed~via the corresponding Ricci tensors
and Ricci scalars.~This means that all curvature tensors can be constructed out of $\eta_{ab}$ 
and $R_{ab}$ only.~However,~$R_{ab} \propto \eta_{ab}$~in (A)dS, whereas
$R_{ab} \propto \delta_{ab} - \eta_{ab}$ in ESUs.~Hence,~in the former case,~$\eta_{ab}K^ay^b$
and $\eta_{ab}y^ay^b$ are the only (simplest) $y$-dependent scalars, 
whereas,~in the latter case,~one of~$\eta_{ab}K^ay^b$, $\eta_{ab}y^ay^b$, 
$R_{ab}K^ay^b$ and $R_{ab}y^ay^b$ functionally depends on the rest.

\section{Covariant solutions}

The derivation of $\textrm{sol}_K^{(\alpha)}(y)$ reduces to solving
\eqref{eq:cfeq} for $\phi^{(\alpha)}(v)$.~First,~by making use of
\beqa\label{eq:D-and-commutator}
\big[D_z^{(n)},\partial_z^2\big] &=& \frac{n(1-n)}{{\sinh^2}{z}}\,D_z^{(n)}
\quad \textrm{with} \quad
D_z^{(n)} \;\equiv\; {\sinh^n}{z}\,  \frac{d^n}{d ({\cosh}{z})^n}\,,
\eeqa
where $n \in \mathbb{N}_0$,~we find that~\eqref{eq:cfeq} can be simplified to
\beqa\label{eq:cfeq-m}
\left(
D_\eta^{(\gamma)}D_\zeta^{(\alpha)}\big(\partial_\eta^2 - \partial_\zeta^2\big)\frac{1}{D_\eta^{(\gamma)}D_\zeta^{(\alpha)}}
+ \frac{(\nu^2 - \mu^2)\big(\partial_\chi^2 + \xi \partial_\xi^2 +1\big)}{{\sinh^2}{\eta}}
\right) \phi^{(\alpha)}(v) &=& 0\,,
\eeqa
where we have analytically continued $n$ in~\eqref{eq:D-and-commutator} to complex numbers.~Second,~we
have~from \eqref{eq:van-var} that $\phi^{(\alpha)}(v)$ is independent of $\xi$.~Third,
$\chi$ and $(\nu^2 - \mu^2)$ vanish in (A)dS.~However,~$\chi$ is the only variable
depending~(linearly)~on the time coordinate~in
ESUs.~We~then~assume~that $\phi^{(\alpha)}(v)$ is an eigenfunction of $\partial_\chi$
with the eigenvalues $\pm i$.~At~last,~making use of the method~of
separation of
variables: $\phi^{(\alpha)}(v) \to \phi_{\pm k}^{(\alpha)}(v)$, defined by
\beqa
\phi_{\pm k}^{(\alpha)}(v) &=& \varphi_{\pm k}^{(\gamma)}(\eta)\, \psi_{k}^{(\alpha)}(\zeta)\,\vartheta(\chi) \,,
\eeqa
where $k \in \mathbb{C}$ is an arbitrary parameter, we obtain from~\eqref{eq:cfeq-m} that
\bsubeqs
\beqa
\frac{d^2}{d\eta^2}\,\frac{1}{D_\eta^{(\gamma)}}\,\varphi_{k}^{(\gamma)}(\eta)&=& +k^2
\frac{1}{D_\eta^{(\gamma)}}\,\varphi_{k}^{(\gamma)}(\eta)\,,
\\[2mm]
\frac{d^2}{d\zeta^2}\,\frac{1}{D_\zeta^{(\alpha)}}\,\psi_{k}^{(\alpha)}(\zeta)&=& +k^2
\frac{1}{D_\zeta^{(\alpha)}}\,\psi_{k}^{(\alpha)}(\zeta)\,,
\\[1mm]
\frac{d^2}{d\chi^2}\,\vartheta(\chi) &=& -\vartheta(\chi)\,.
\eeqa
\esubeqs
A particular solution of these equations we wish to consider is given by
\bsubeqs\label{eq:sep-sol}
\beqa\label{eq:sep-sol-1}
\varphi_{k}^{(\gamma)}(\eta) &\equiv&
\frac{{\Gamma}{(1 -\gamma -k)}\,
{\Gamma}{(\gamma -k)}}{\left(\scalebox{0.9}{$i\sqrt{\gamma(1-\gamma)}$}\right)^{-k}\Gamma(\scalebox{0.9}{$1-k$})}\,e^{k\eta}
{{}_2F_1}{\left(\scalebox{0.9}{$\gamma$},\scalebox{0.9}{$1-\gamma$},\scalebox{0.9}{$1-k$};\scalebox{0.9}{$\dfrac{1}{1-e^{2\eta}}$}\right)},
\\[0.5mm]\label{eq:sep-sol-2}
\psi_{k}^{(\alpha)}(\zeta) &\equiv& (-1)^\alpha\,
\frac{\Gamma(\alpha - k)}{\Gamma(-k)}\,e^{k \zeta}
{{}_2F_1}{\left(\scalebox{0.9}{$\alpha$},\scalebox{0.9}{$1-\alpha$},\scalebox{0.9}{$1-k$};
\scalebox{0.9}{$\dfrac{1}{1-e^{2\zeta}}$}\right)},
\\[3mm]\label{eq:sep-sol-3}
\vartheta(\chi) &\equiv& e^{i\chi}\,,
\eeqa
where
$\Gamma(z)$ and ${{}_2F_1}(a,b,c;z)$ are,~respectively,~the gamma and hypergeometric functions.~The
coefficients independent of $\eta$ and $\zeta$ have been chosen
in~\eqref{eq:sep-sol-1} and~\eqref{eq:sep-sol-2} in order
to simplify certain expressions which appear below (see also Sec.~III\,C\,1 in~\cite{Emelyanov-2020}).~Besides,~it
proves useful to consider
\beqa\label{eq:sep-sol-4}
\tilde{\varphi}_{k}^{(\gamma)}(\eta) &\equiv&
\frac{\Gamma(\scalebox{0.9}{$k$})}{\left(\scalebox{0.9}{$i\sqrt{\gamma(1-\gamma)}$}\right)^{k}}\,e^{-k\eta}
{{}_2F_1}{\left(\scalebox{0.9}{$\gamma$},\scalebox{0.9}{$1-\gamma$},\scalebox{0.9}{$1-k$};\scalebox{0.9}{$\dfrac{1}{1-e^{-2\eta}}$}\right)},
\eeqa
\esubeqs
which is related to~\eqref{eq:sep-sol-1} as follows:
\beqa\label{eq:sep-sol-1-via-sep-sol-4}
\varphi_k^{(\gamma)}(\eta) &=& \frac{\Gamma(\scalebox{0.9}{$1-\gamma-k$})\,\Gamma(\scalebox{0.9}{$\gamma-k$})}{\Gamma(\scalebox{0.9}{$\gamma$})\,\Gamma(\scalebox{0.9}{$1-\gamma$})
\left(\scalebox{0.9}{$\gamma(1-\gamma)$}\right)^{-k}}\,
\tilde{\varphi}_k^{(\gamma)}(\eta) +
\tilde{\varphi}_{-k}^{(\gamma)}(\eta)\,,
\eeqa
where we have made use of 9.131.2 on p.~1008
in~\cite{Gradshteyn&Ryzhik}.

Note,~\eqref{eq:sep-sol} can be obtained by getting rid of the $D$-operator via~\eqref{eq:D-and-commutator},
or by observing
that
\beqa\label{eq:psi-via-D}
\psi_{k}^{(\alpha)}(\zeta) &=& D_\zeta^{(\alpha)} e^{k\zeta}\,,
\eeqa
where~an integral representation of this fractional-derivative operator can be readily deduced by 
using the Fourier transform and one of the integral representations of the hypergeometric
function.

We have,~thus,~established by construction that there exist
non-perturbative solutions of the Klein-Gordon
equation being scalars with respect to general coordinate transformations: 
\beqa
\phi^{(\alpha)}(v) &=& {\int}{dk}\, \big(c_{k}^{(+)} \phi_{+ k}^{(\alpha)}(v) 
+ c_{k}^{(-)} \phi_{- k}^{(\alpha)}(v)\big)\,,
\eeqa
where the integration over $k \in \mathbb{C}$ and the coefficients $c_{k}^{(\pm)}$
need to be determined  
on
physical grounds.

\section{Wightman functions}

A Wightman 2-point function
\beqa
W(\sigma) &\equiv& \langle\Omega| \hat{\Phi}(x)\hat{\Phi}(X) | \Omega\rangle\,
\eeqa
describes quantum-field correlations,~see Sec.~6.6 in~\cite{Mukhanov&Winitzki} for further details.~The Wightman 
functions for all spacetimes under
study have been obtained~so~far~in~the literature by making 
use of the non-covariant approach to quantum particle
physics~\cite{Birrell&Davies}.~In this section,~we~intend
to determine their
momentum-space representations with the Lorentz-invariant 
momentum-integral measure
as in Minkowski spacetime~\cite{Bunch&Parker}:
\beqa\label{eq:wf}
W^{(\alpha)}(\sigma) &=&{\int}\frac{d^dK}{(2\pi)^{d-1}}\,\theta\big(K^0\big)\,\delta\big(\eta_{ab}K^aK^b - M^2\big)\,\textrm{sol}_K^{(\alpha)}(y)\,,
\eeqa
where $\theta(z)$ is the Heaviside function.

\subsection{Anti-de-Sitter and de-Sitter universes}
\label{sec:wf-ds}

\begin{figure}
\includegraphics[scale=0.65]{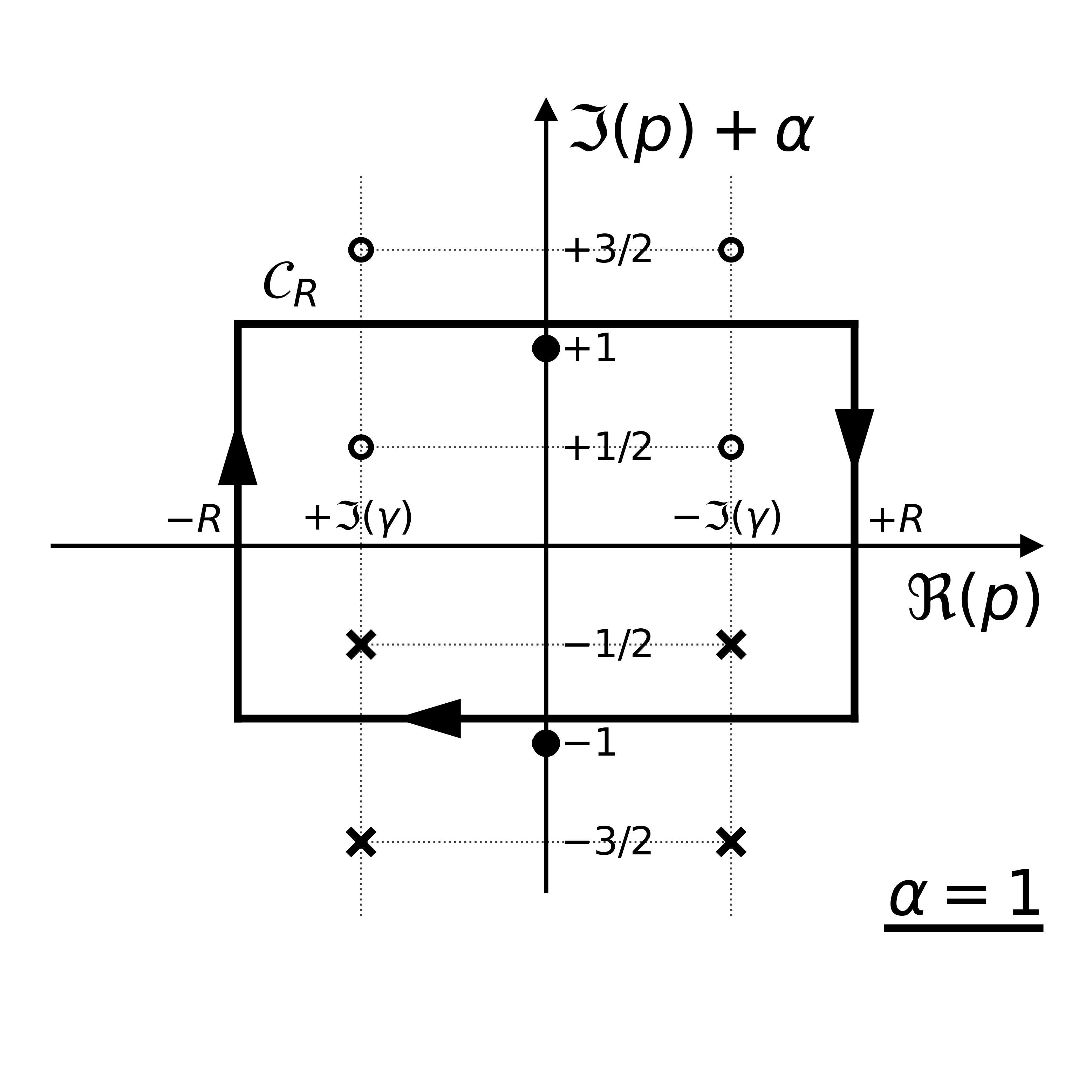}
\includegraphics[scale=0.65]{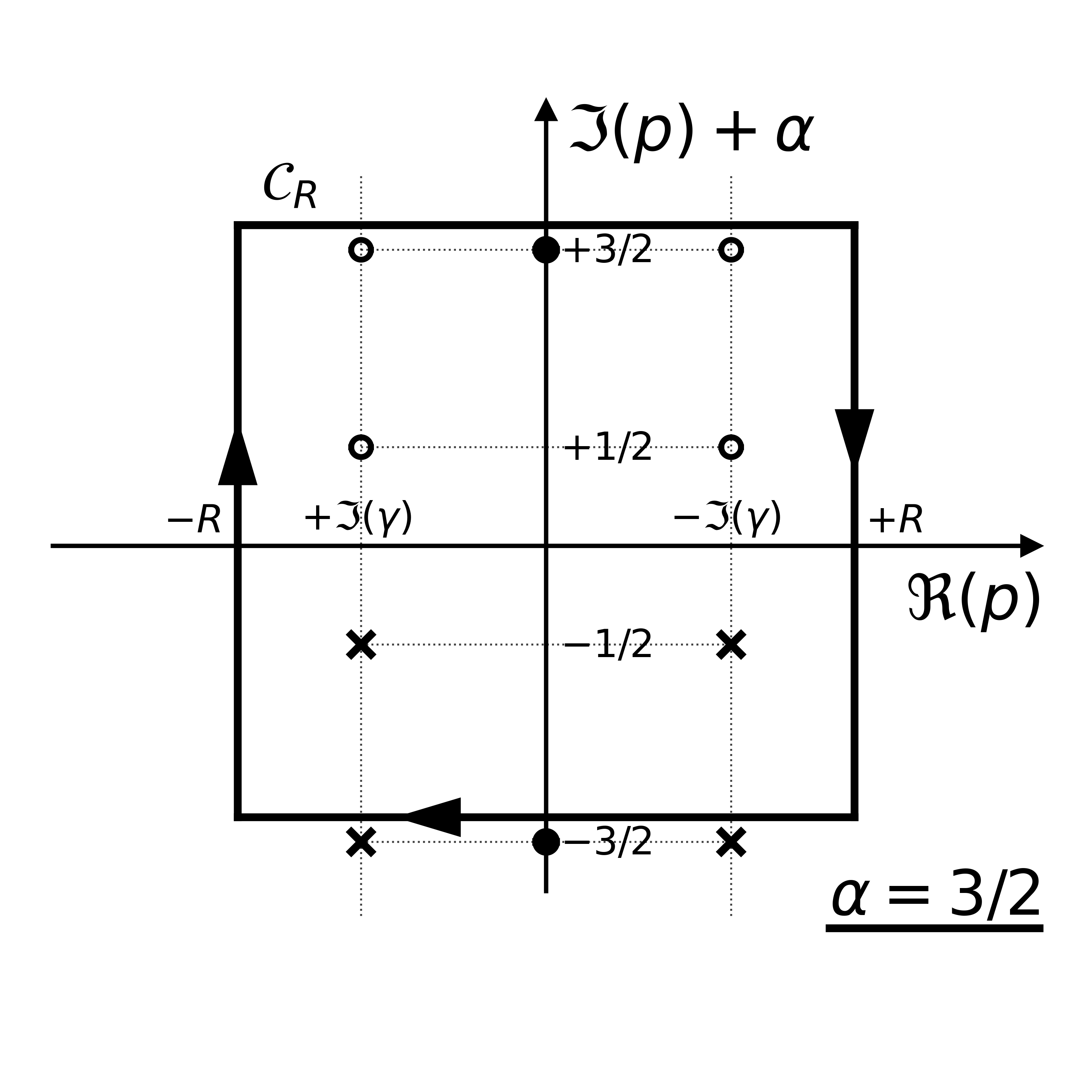}\vspace{-10mm}
\caption{A complex $p$-plane with poles in the integrand of~\eqref{eq:phi-i}
shown by cross and empty-dot~marks originating from $\varphi_{ip-\alpha}^{(\gamma)}(\eta)$
and $\varphi_{\alpha-ip}^{(\gamma)}(\eta)$,~respectively.~After the integration over
momentum, a pair of extra first-order poles at $p = 0$ and $p = -2i\alpha$ emerge,~being~marked
by solid dots.~We~then~choose a rectangular contour with the lower side at
$\Im(p) = -2\alpha + 0$ in order to evaluate the integral~over~$p$.
Left:~For the integer
values of $\alpha$, e.g. $\alpha = 1$,~the cross and empty-dot poles lying
on a line~to~pass through $p = -i\alpha$ give residues which cancel each other.~Right:~For
the half-integer values of $\alpha$,~e.g. $\alpha = \frac{3}{2}$, we set
$c_{\alpha - ip}^{(\alpha)} = 0$ at $p = \pm \Im(\gamma)$ to avoid the residues at these poles.}\label{fig:poles-2}
\end{figure}

Generalising the results of Sec.~3.3.1 in~\cite{Emelyanov-2020} to $\alpha \neq 1$,~we
consider
\bsubeqs
\beqa\label{eq:phi-i}
\phi_{1}^{(\alpha)}(v) &\equiv& e^{i\chi}
{\int_\mathcal{C}}{dp}\,\psi_{ip-\alpha}^{(\alpha)}(\zeta)
\Big(c_{ip-\alpha}^{(\alpha)}\,\varphi_{ip-\alpha}^{(\gamma)}(\eta) + 
c_{\alpha - ip}^{(\alpha)}\,\varphi_{\alpha-ip}^{(\gamma)}(\eta)\Big),
\eeqa
where the contour $\mathcal{C}$ corresponds to the integration
over $\Re(p) \in (-\infty,+\infty)$ with $\Im(p) = +i0$,
because then we have in the limit of vanishing 
curvature~($C \to 0$)~that
\beqa\label{eq:phi-i-ctzl}
\textrm{sol}_{K,1}^{(\alpha)}(y) &\equiv&  \frac{1}{4\pi}\,\frac{{\sinh^\alpha}{\eta}}{{\sinh^\alpha}{\zeta}}\,
\frac{\phi_{1}^{(\alpha)}(v)}{\left(\scalebox{0.9}{$i\sqrt{\gamma(1-\gamma)}$}\right)^{\alpha}}\,
\,\xrightarrow[C \,\to\, 0]{}\,
\exp_K(y)\,,
\eeqa
where we have assumed
\beqa\label{eq:c-con-1}
c_{+k}^{(\alpha)}\big|_{C \,\to\, 0} + c_{-k}^{(\alpha)}\big|_{C \,\to\, 0} &=& 2\,.
\eeqa
\esubeqs
Hence,~\eqref{eq:phi-i} plugged into~\eqref{eq:c-phi} yields a positive-frequency
plane wave in the limit~of vanishing curvature.~A superposition
of such plane waves with various on-shell values~of~$K$ gives~a~wave packet
which adequately models particles in collider physics,~as explained~in~Sec.~\ref{sec:pm}.

Plugging~\eqref{eq:phi-i-ctzl} with~\eqref{eq:phi-i} into~\eqref{eq:wf},~we first integrate
over $K$.~This 
yields a result~which generalises (47) in~\cite{Emelyanov-2020} to $\alpha \neq 1$.~We then integrate
over $p$ by use of the residue~theorem,~see fig.~\ref{fig:poles-2} for more details.~We obtain this way that
\bsubeqs
\beqa\label{eq:wf-i}
W_{1}^{(\alpha)}(\sigma) &=& \frac{{\sinh^\alpha}{\eta}}{(4\pi)^{1+\alpha}}\,
\frac{(-2)^\alpha M^{2\alpha}}{2}\,
\frac{
c_{-\alpha}^{(\alpha)}\,\varphi_{-\alpha}^{(\gamma)}(\eta) + 
c_{+\alpha}^{(\alpha)}\,\varphi_{+\alpha}^{(\gamma)}(\eta)}{\left(\scalebox{0.9}{$i\sqrt{\gamma(1-\gamma)}$}\right)^{\alpha}}\,,
\eeqa
where we have assumed for half-integer values of $\alpha$ that
\beqa\label{eq:c-con-2}
c_{\alpha \pm i\Im(\gamma)}^{(\alpha)}\big|_{\alpha + \frac{1}{2} \,\in\, \mathbb{N}} &=& 0\,,
\eeqa
\esubeqs
to avoid extra contributions to~\eqref{eq:wf-i},~as outlined in~fig.~\ref{fig:poles-2}.~By making use of~$\varphi_{+\alpha}^{(\gamma)}(\eta) \propto \varphi_{-\alpha}^{(\gamma)}(\eta)$
for $\alpha \in \mathbb{N}_0$,
which follows from the first formula~on 
p.~38 in \cite{Magnus&Oberhettinger&Soni}, we also assume
\bsubeqs
\beqa\label{eq:c-ass-1}
c_{-\alpha}^{(\alpha)} + c_{+\alpha}^{(\alpha)}\,
\frac{(\gamma)_{\alpha} (1-\gamma)_{\alpha}}
{\gamma^{\alpha}(1-\gamma)^{\alpha}}
&=& 2\,,
\\[1mm]\label{eq:c-ass-2}
c_{+\alpha}^{(\alpha)}\big|_{\alpha + \frac{1}{2} \,\in\, \mathbb{N}} &=& 0\,,
\eeqa
\esubeqs
where $(z)_\nu$ is the Pochhammer function.~These conditions on the $c$-coefficients reduce~\eqref{eq:wf-i}~to
\beqa
W_{1}^{(\alpha)}(\sigma) &=&
\frac{H^{2\alpha}\Gamma(\scalebox{0.9}{$\gamma +\alpha$})\Gamma(\scalebox{0.9}{$1-\gamma + \alpha$})}{(4\pi)^{1+\alpha}\,\Gamma(\scalebox{0.9}{$1+\alpha$})}
\,{{}_2F_1}{\left(\scalebox{0.9}{$\gamma +\alpha$},\scalebox{0.9}{$1-\gamma + \alpha$},\scalebox{0.9}{$1+\alpha$};
\scalebox{0.9}{$\dfrac{1+{\cosh}{\sqrt{2H^2\sigma}}}{2}$}\right)}.
\eeqa
This is the Wightman function in the Chernikov-Tagirov~\cite{Chernikov&Tagirov} or
Bunch-Davies~\cite{Bunch&Davies} state~in~dS, where $H$ is a Hubble parameter of de-Sitter spacetime,~see e.g.~\cite{Birrell&Davies,Akhmedov}.~In the
case of AdS,~we analytically continue $H$ to $iH$ in the Wightman function, see~\cite{Avis&Isham&Storey}.

\subsection{Closed and open Einstein static universes}
\label{sec:wf-esu}

\begin{figure}
\includegraphics[scale=0.65]{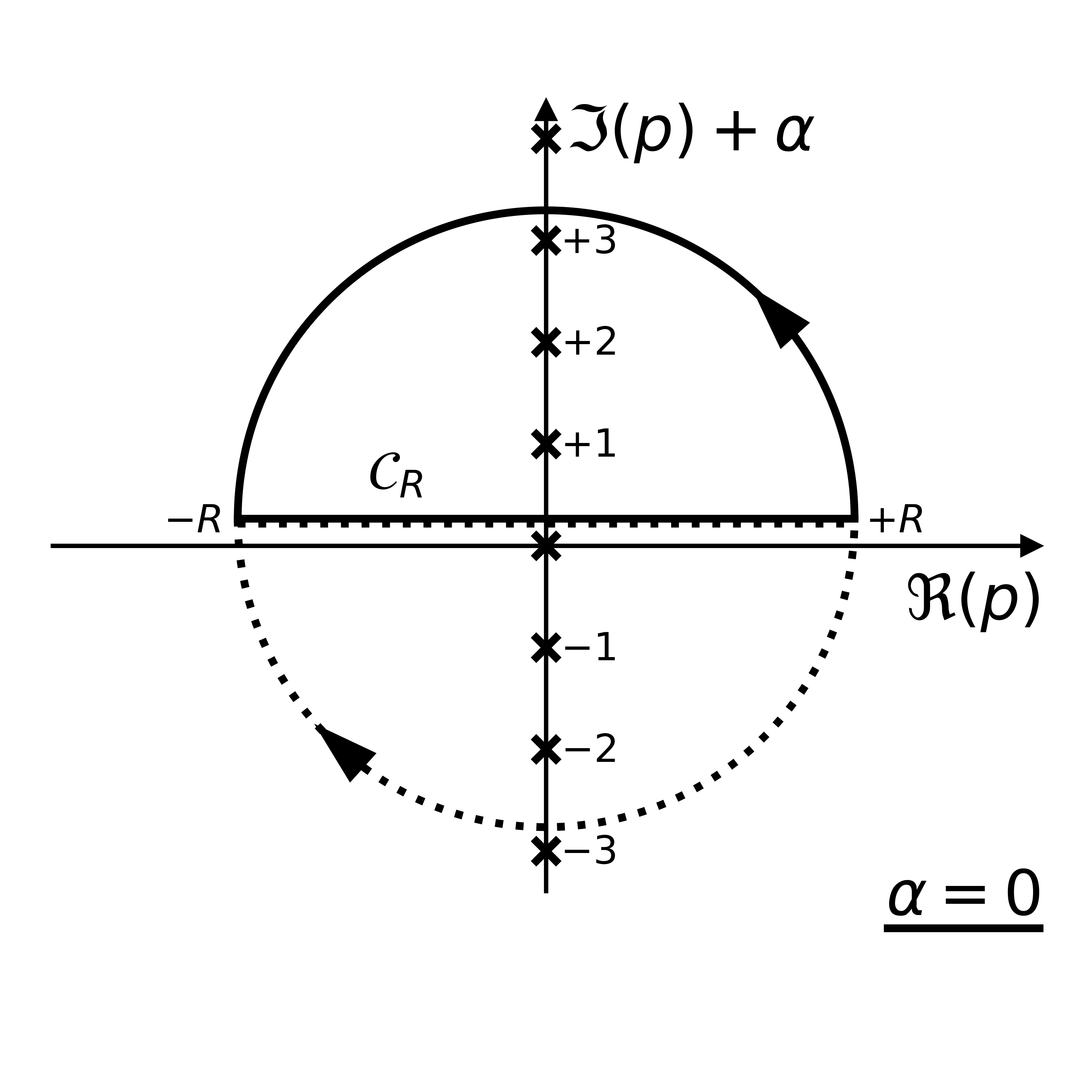}
\includegraphics[scale=0.65]{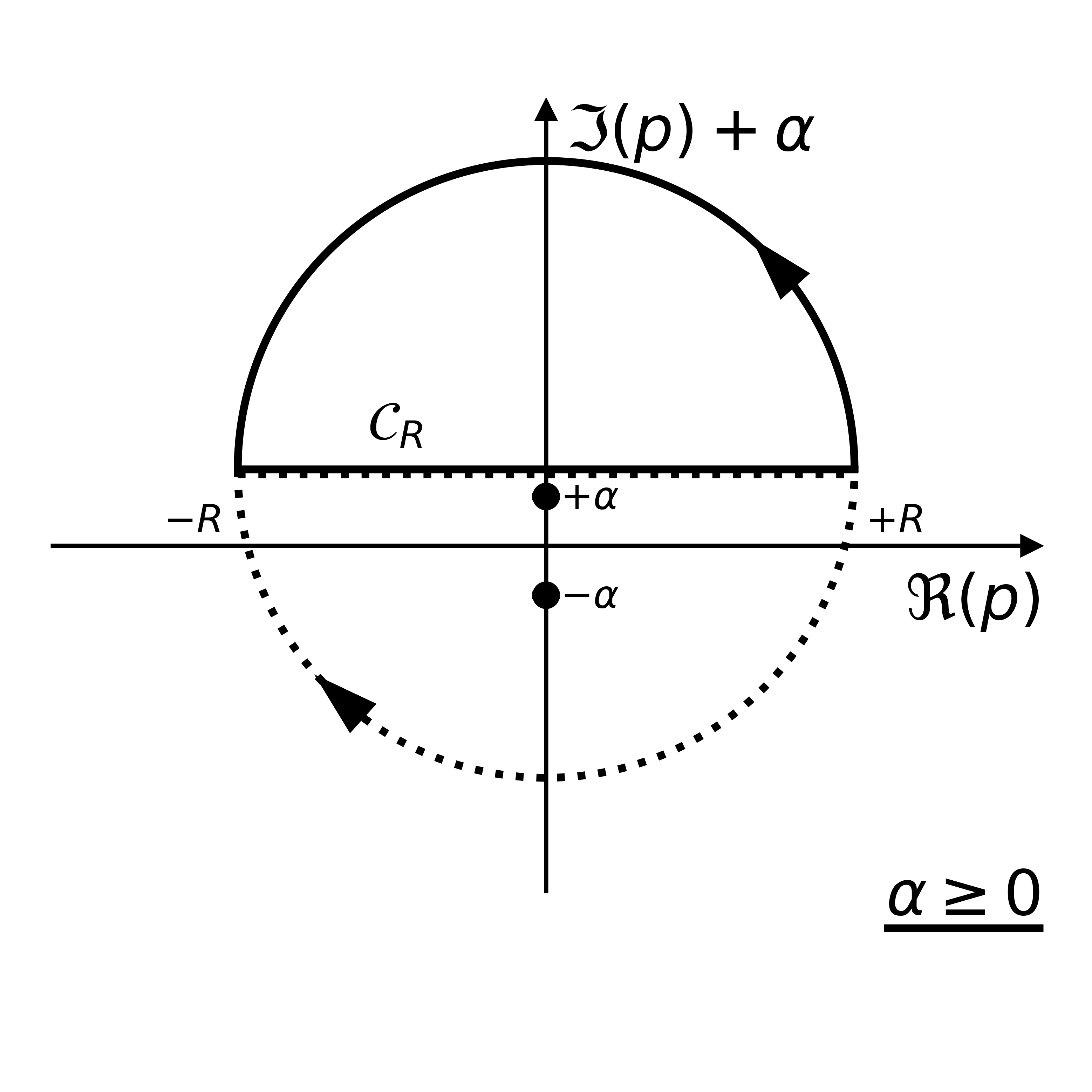}\vspace{-10mm}
\caption{Left:~A complex $p$-plane with poles in the integrand of~\eqref{eq:phi-ii} shown
by cross marks.~By~use of the residue theorem, the solid contour is chosen
for the integrand part involving $\tilde{\varphi}_{ip-\alpha}^{(\gamma)}(\eta)$ as this
vanishes for $\Im(p) > 0$ faster than any exponential function in the limit $R \to \infty$.~For this property~to
hold~for~the
integrand part in~\eqref{eq:phi-ii} involving
$\tilde{\varphi}_{\alpha - ip}^{(\gamma)}(\eta)$, the dotted contour is used.~Right:~The pole
structure~of \eqref{eq:phi-ii} alters after the integration over the angles in
momentum space.~The cross-marked first-order poles disappear,~while solid-dot-marked
first-order poles emerge at $p = 0$ and $p = -2i\alpha$,
which yield the result~\eqref{eq:wf-ii} by making use of the residue theorem.}\label{fig:poles-3}
\end{figure}

We now wish to consider a particular solution of the covariant scalar-field equation~\eqref{eq:cfeq}~of
the following form:
\bsubeqs
\beqa\label{eq:phi-ii}
\phi_{2}^{(\alpha)}(v) &\equiv& {e^{i\chi}}
{\int_\mathcal{C}}{dp}\,
\Big(\psi_{ip-\alpha}^{(\alpha)}(\zeta) + \psi_{\alpha-ip}^{(\alpha)}(\zeta)\Big)
\Big(
\tilde{c}_{ip-\alpha}^{(\alpha)}\,\tilde{\varphi}_{ip-\alpha}^{(\gamma)}(\eta) +
\tilde{c}_{\alpha - ip}^{(\alpha)}\,\tilde{\varphi}_{\alpha-ip}^{(\gamma)}(\eta)\Big),
\eeqa
such that
\beqa\label{eq:phi-ii-ctzl}
\textrm{sol}_{K,2}^{(\alpha)}(y) &\equiv&  \frac{1}{4\pi}\,\frac{{\sinh^\alpha}{\eta}}{{\sinh^\alpha}{\zeta}}\,
\frac{\phi_{2}^{(\alpha)}(v)}{\left(\scalebox{0.9}{$i\sqrt{\gamma(1-\gamma)}$}\right)^{\alpha}}\,
\,\xrightarrow[C \,\to\, 0]{}\,
\exp_K(y)\,,
\eeqa
with the assumption
\beqa\label{eq:tc-con-1}
\tilde{c}_{n \,\in\,\mathbb{Z}}^{(\alpha)}\big|_{C\,\to\, 0} &=& 1\,.
\eeqa
\esubeqs
In fact, we have from~10.16.10 on p.~228 in~\cite{Olver&etal} that
\beqa
\tilde{\varphi}_{k}^{(\gamma)}(\eta)  &\xrightarrow[C \,\to\, 0]{}& \frac{\pi}{{\sin}(\pi k)}\,e^{-i\pi k/2}
J_{-k}\Big(2\sqrt{\gamma(1-\gamma)}e^\eta\Big)\,,
\eeqa
where $J_\nu(z)$ is the Bessel function,~cf.~(38) in~\cite{Emelyanov-2020}.~Thus,~there
are poles at
$ip - \alpha  \in \mathbb{Z}$,~which need to be taken into account by integrating
over $p$ in~\eqref{eq:phi-ii},~see fig.~\ref{fig:poles-3},~left,~for more details.
The residue theorem and
5.7.5.2 on p.~584 in~\cite{Prudnikov&etal} give~\eqref{eq:phi-ii-ctzl},~assuming~\eqref{eq:tc-con-1} holds.

Plugging~\eqref{eq:phi-ii-ctzl} with~\eqref{eq:phi-ii} into~\eqref{eq:wf},~we first introduce
spherical coordinates
in momentum space.~We then integrate over the angles which gives rise to poles at $p = 0$ and
$p = -2i\alpha$ in the complex $p$-plane, as shown in fig.~\ref{fig:poles-3},~right.~This gives
by use of the residue theorem that
\beqa\label{eq:wf-ii}
W_{2}^{(\alpha)}(\sigma) &=& \frac{{\sinh^\alpha}{\eta}}{(4\pi)^{1+\alpha}}\,\frac{(-2)^\alpha}{2}
{\int\limits_0^{+\infty}}\frac{dK K^{2\alpha+1}e^{i\chi}}{\sqrt{M^2+K^2}}\,
\frac{\tilde{c}_{-\alpha}^{(\alpha)}\,\tilde{\varphi}_{-\alpha}^{(\gamma)}(\eta) - 
\tilde{c}_{+\alpha}^{(\alpha)}\,\tilde{\varphi}_{+\alpha}^{(\gamma)}(\eta)}
{\left(\scalebox{0.9}{$i\sqrt{\gamma(1-\gamma)}$}\right)^{\alpha}\Gamma(\scalebox{0.9}{$-\alpha$})
\Gamma(\scalebox{0.9}{$1+\alpha$})}\,.
\eeqa
We next assume
\bsubeqs
\beqa\label{eq:tc-ass-1}
\tilde{c}_{-\alpha}^{(\alpha)} + \tilde{c}_{+\alpha}^{(\alpha)}\,
\frac{(\gamma)_{\alpha} (1-\gamma)_{\alpha}}
{\gamma^{\alpha}(1-\gamma)^{\alpha}}
&=& 2\,,
\\[1mm]\label{eq:tc-ass-2}
\tilde{c}_{+\alpha}^{(\alpha)}\big|_{\alpha + \frac{1}{2} \,\in\, \mathbb{N}} &=& 0\,,
\eeqa
\esubeqs
because then the integrand in~\eqref{eq:wf-ii} reduces (up to a factor) to
the results given in (21) and~in (32) in the reference~\cite{Bander&Itzykson}.~This can be shown
by successively using 9.131.1(3), 9.132.1, 9.134.3, 9.131.1(1), 9.131.1(3) in~\cite{Gradshteyn&Ryzhik}.~Therefore,
we finally obtain
\beqa\label{eq:wf-ii-m}
W_{2}^{(\alpha)}(\sigma) &=& \frac{(1-e^{2\eta})^{\alpha}}{(4\pi)^{1+\alpha}\Gamma(\scalebox{0.9}{$1+\alpha$})}
{\int\limits_0^{+\infty}}\frac{dK K^{2\alpha+1}e^{i\chi}}{\sqrt{M^2+K^2}}\,
{{}_2F_1}{\left(\scalebox{0.9}{$\gamma$},\scalebox{0.9}{$1-\gamma$},\scalebox{0.9}{$1+\alpha$};\scalebox{0.9}{$\dfrac{1}{1-e^{-2\eta}}$}\right)}.
\eeqa

Taking into account $\gamma = 1/2 + Ka$ in $\textrm{CESU}_4$,~where $a$ is the radius of
the~three-dimensional sphere -- spatial section of $\textrm{CESU}_4$, -- we find by making use of 9.121.16 on p.~1006 in~\cite{Gradshteyn&Ryzhik}~with analytic continuation to non-integer values of $2Ka$ that
\beqa\label{eq:wf-ii-1/2}
W_{2}^{(\frac{1}{2})}(\sigma) &=&
\frac{|\boldsymbol{y}|/a}{\sin(|\boldsymbol{y}|/a)}\,
\frac{M^2K_1\hspace{-0.75mm}\left(\sqrt{-2M^2\sigma}\right)}{4\pi^2\sqrt{-2M^2\sigma}}
\bigg|_{|\boldsymbol{y}| \,\equiv\, a\sqrt{\frac{1}{2}R_{ab}\partial^a\sigma \partial^b\sigma}}\,,
\eeqa
where $K_\nu(z)$ is the modified Bessel function.~This result fits to that found
in~\cite{Dowker&Critchley}~if~we~replace $|\boldsymbol{y}|$ by
$|\boldsymbol{y}| + 2\pi a n$ and then sum over $n \in \mathbb{Z}$.~It should be mentioned
that \eqref{eq:wf-ii-m} is periodic~on~the sphere (or continuous over all great circles),~as
$\eta$ is oblivious to that replacement.~Still,~\eqref{eq:wf-ii-1/2}~is non-periodic due to
the analytic continuation used above.~In the case~of $\textrm{OESU}_4$,~we
reproduce the result of~\cite{Bunch}~by analytically~continuing~$a$~to~$ia$~in~\eqref{eq:wf-ii-1/2}.

\section{A single covariant solution}
\label{sec:single-cs}

We have found the solutions
$\textrm{sol}_{K,1}^{(\alpha)}(y)$~and~$\textrm{sol}_{K,2}^{(\alpha)}(y)$
in the last section.~The~former~has~been
shown to be linked through~\eqref{eq:wf} with the Wightman function in dS,~while
the latter~to~that~in ESU.~We wish now to study if it is possible to have a \emph{single}
solution~$\textrm{sol}_{K}^{(\alpha)}(y)$~being 
physically acceptable -- plane wave as $C \to 0$ -- in all spacetimes under consideration.~The
strategy~is~to compare $\textrm{sol}_{K,1}^{(\alpha)}(y)$ with $\textrm{sol}_{K,2}^{(\alpha)}(y)$ in ESU,
as both of these approach $\exp_K(y)$ at $C \to 0$.

Both $\textrm{sol}_{K,1}^{(\alpha)}(y)$~and~$\textrm{sol}_{K,2}^{(\alpha)}(y)$
contain unknown coefficients,~$c_{\pm ip \mp\alpha}^{(\alpha)}$ and
$\tilde{c}_{\pm ip \mp\alpha}^{(\alpha)}$,~see~\eqref{eq:phi-i}~and
\eqref{eq:phi-ii}.~We~wish to focus on a particular case which may be of interest in physics.~Specifically,
we focus here on
$\text{dS}_d$ and $\text{CESU}_{d+1}$ with $d \in \{2,4\}$,~or,~in other words, $\alpha \in \{0,1\}$,~such~that\;\;
\bsubeqs
\beqa\label{eq:phi-i-s}
\phi_{1}^{(\alpha)}(v) &=& e^{i\chi}{\int_\mathcal{C}}{dp}
\,\psi_{ip-\alpha}^{(\alpha)}(\zeta)
\Big(\varphi_{ip-\alpha}^{(\gamma)}(\eta) + \varphi_{\alpha-ip}^{(\gamma)}(\eta)\Big),
\\[1mm]\label{eq:phi-ii-s}
\phi_{2}^{(\alpha)}(v) &=& 
{e^{i\chi}}{\int_\mathcal{C}}{dp}\,
\Big(\psi_{ip-\alpha}^{(\alpha)}(\zeta) + \psi_{\alpha-ip}^{(\alpha)}(\zeta)\Big)
\Big(\tilde{\varphi}_{ip-\alpha}^{(\gamma)}(\eta) +
\tilde{\varphi}_{\alpha-ip}^{(\gamma)}(\eta)\Big).
\eeqa
\esubeqs
The $c$- and $\tilde{c}$-coefficients have thus been set to unity,~cf. (35) and (36) in~\cite{Emelyanov-2020}.~This agrees~with all conditions imposed above on these 
coefficients.~In this case,~we
can evaluate the integrals over $p$, namely we obtain from~\eqref{eq:phi-i-s} and~\eqref{eq:phi-ii-s} that
\bsubeqs
\beqa\label{eq:phi-a}
\phi_{1,2}^{(\alpha)}(v) &=& 2\pi
{e^{i\chi}}\sum\limits_{l \,=\, 0}^{+\infty} \frac{(\gamma)_l (1-\gamma)_l}{l! (1-e^{-2\eta})^l}
\sum\limits_{n \,\in\, S_{1,2}(l)}
\frac{\psi_{+n}^{(\alpha)}(\zeta)+\psi_{-n}^{(\alpha)}(\zeta)}{(-e^{-\eta})^{n}\Gamma(\scalebox{0.9}{$n + l+1$})}\,
(\gamma(\gamma-1))^\frac{n}{2}\,,
\eeqa
where
\beqa\label{eq:phi-a-i}
S_1(l) &\equiv& \mathbb{Z}_{\,\geq\, -l}\,, 
\\[1mm]\label{eq:phi-a-ii}
S_2(l) &\equiv& \mathbb{N}_{\,\geq\, 1+\alpha}\uplus\mathbb{Z}_{\,\geq\, -\alpha}\,.
\eeqa
\esubeqs
The right-hand side
of~\eqref{eq:phi-a} with~\eqref{eq:phi-a-i} can~be summed over both $n$ and $l$.~This yields~\eqref{eq:phi-a} which
agrees with (42) and (43) from Sec.~III\,C\,1
in~\cite{Emelyanov-2020} in the case of $\alpha = 1$.~This case~can~be straightforwardly extended
to $\alpha = 0$.~The sum over~$l$~in \eqref{eq:phi-a} with~\eqref{eq:phi-a-ii} is a definition~of~the
ordinary hypergeometric function,~whereas the sum over $n$ follows from the~residue-theorem
application in the complex $p$-plane.~Specifically,~we have from
\beqa
\tilde{\varphi}_{k}^{(\gamma)}(\eta) &=& 
\sum\limits_{l \,=\, 0}^{+\infty} \frac{(\gamma)_l (1-\gamma)_l}{l! (1-e^{-2\eta})^l}\,
(-1)^l \,\Gamma(k-l)\,e^{-k(\eta + \ln i\sqrt{\gamma(1-\gamma)})}\,
\eeqa
that there are poles at $p = -i (\alpha \mp m \pm l)$ for $k = \pm (ip-\alpha)$, where $m \in \mathbb{N}_0$.~By the
integration over $p$ in~\eqref{eq:phi-ii-s},~we take the poles with $m \geq   l + \alpha + 1$ for the semi-circle contour in
the upper complex half-$p$-plane,~while $m \geq l -\alpha$ in the lower one,~see
fig.~\ref{fig:poles-3},~left.~The~residues at these poles give in the end rise to the summation with respect to
$n$ over the multiset~\eqref{eq:phi-a-ii}.

No free parameters enter $\textrm{sol}_{K,1}^{(\alpha)}(y)$~and~$\textrm{sol}_{K,2}^{(\alpha)}(y)$
for $\alpha \in \{0,1\}$.~We compare~these~solutions 
with respect~to the 
difference of the corresponding Wightman functions in ESU.~According to~\eqref{eq:wf}, we particularly need 
to integrate over the angles in momentum space.~The non-trivial part of this 
integration comes from
the variable $\zeta$ as this is the only variable depending~on~the angles through
$\boldsymbol{K}{\cdot}\boldsymbol{y}$.~Assuming
$\boldsymbol{K}{\cdot}\boldsymbol{y} = |\boldsymbol{K}||\boldsymbol{y}|\cos\theta$,~we obtain
\beqa
\zeta &=& {\ln}{\big({\cos\theta} + \sqrt{\cos^2\theta - 1}\big)}\,. 
\eeqa
This means that the right-hand side of
\beqa\label{eq:w1-w2}
W_1^{(\alpha)}(\sigma) - W_2^{(\alpha)}(\sigma) &\propto&
{\int}{\frac{d^{d-1}\boldsymbol{K}}{\sqrt{M^2 + \boldsymbol{K}^2}}}\,\Big(\textrm{sol}_{K,1}^{(\alpha)}(y) - \textrm{sol}_{K,2}^{(\alpha)}(y)\Big)
\eeqa
has the following integration part:
 \beqa\label{eq:w1-w2-p}
{\int\limits_0^{(2-\alpha)\pi}}\hspace{-2.5mm}d\theta\,{\sin^{2\alpha}}{\theta}\,
\frac{\phi_{1}^{(\alpha)}(v)-\phi_{2}^{(\alpha)}(v)}
{{\sinh^\alpha}{(\zeta)}}
&\propto& 
{\sum\limits_{l \,=\, 0}^{+\infty} \frac{(\gamma)_l (1-\gamma)_l}{l! (1-e^{-2\eta})^l}}
\sum\limits_{n \,\in\, S_{1}(l){\triangle}S_{2}(l)}
{\frac{\delta_{n,+\alpha} + \delta_{n,-\alpha}}{\Gamma(\scalebox{0.9}{$n + l+1$})}}\,.
\eeqa
The right-hand side of~\eqref{eq:w1-w2-p} vanishes if $\alpha = 0$ as $0 \notin S_{1}(l){\triangle}S_{2}(l)$ for any 
$l \geq 0$,~where~$\triangle$~stands for the symmetric difference of the sets.\,This~also~vanishes
if $\alpha = 1$,~because~$\pm 1 \notin S_{1}(l){\triangle}S_{2}(l)$ for $l \geq 1$
and the contribution from $- 1 \in S_{1}(0){\triangle}S_{2}(0)$ gives $1/\Gamma(z) \to 0$ at $z \to 0$.~Therefore,
$W_1^{(\alpha)}(\sigma)$
and $W_{2}^{(\alpha)}(\sigma)$ are equal in ESU for $\alpha \in \{0,1\}$.

Accordingly,~the single solution $\textrm{sol}_{K}^{(\alpha)}(y)$ of \eqref{eq:ofeq} we
have looked
for is given by $\textrm{sol}_{K,1}^{(\alpha)}(y)$.~It is,~generically,~non-unique.~However,
this solution yields the Wightman functions in $\textrm{dS}_{d}$~and $\textrm{CESU}_{d+1}$ with $d \in \{2,4\}$,~which are identified with those previously found in the literature.
Furthermore, by making use of analytic continuation,~we also cover here the spacetimes~$\textrm{AdS}_{d}$ and $\textrm{OESU}_{d+1}$.~Finally,~Minkowski spacetime
is covered by considering the limit $C \to 0$.

\section{Quantum particles in the strong-gravity regime}

Wave packets describing quantum particles are subjected to spreading.~This effect might
occur in its enhanced form for a many-boson system occupying a single state~--~Bose-Einstein
condensate~\cite{Pitaevskii&Stringari}.~Such a quantum system is described by a single wave packet.~Focusing~on~the
center-of-mass motion of a Bose-Einstein condensate,~the question arises how it propagates~in
gravity.~We have shown in~\cite{Emelyanov-2020} that quantum particles move along
non-geodesic trajectories 
if their quantum size is non-negligible with respect to the Hubble length of a dS universe.~It
is,~however,~unfeasible to test this in practice.~In practice,~one may consider~a~Bose-Einstein
condensate trapped into~a~2-dimensional sphere.~Alternatively,~the Bose-Einstein 
condensate may be made of quasi-particles excited on a surface of
a many-particle system of a spherical shape.~By letting it freely move and spread over~the~sphere,~one 
may be able to study~the~non-
perturbative influence of the sphere's curvature on the quantum object.

A 3-dimensional closed Einstein static universe is characterised by the line element
\beqa
ds^2\big|_{\textrm{CESU}_3} &=& dt^2 - a^2\big(d\theta^2 + \sin^2\theta d\phi^2\big)\,,
\eeqa
where $(t,\theta,\phi)$ are linked to $x$ from Sec.~\ref{sec:metric-tensor}
as follows:~$x^0 = t$ and $\boldsymbol{x} = 2a\tan(\theta/2)(\cos\phi,\sin\phi)$.
It is a 2-dimensional sphere with time evolution at each of its points.~It is,~therefore,~tempting
to expect that the center-of-mass motion of a Bose-Einstein condensate on a sphere may be
non-perturbatively 
modelled in theory by the exact solution $\textrm{sol}_{K,1}^{(0)}(y)$ in
$\text{CESU}_3$~from~Sec.~\ref{sec:single-cs}.
It still remains to be clarified to what extent this falls into the setup of
analogue gravity~from Bose-Einstein condensates~\cite{Barcelo&Liberati&Visser}~and~the
description of their center-of-mass motion~\cite{Meister&etal}.

According to~\eqref{eq:psi-via-D},~the solutions $\textrm{sol}_{K,1}^{(0)}(y)$ in $\text{CESU}_3$ and
$\textrm{sol}_{K,1}^{(1)}(y)$ in $\text{CESU}_5$ are~related:\;\;
\beqa\label{eq:alpha1-alpha0}
\textrm{sol}_{K,1}^{(1)}(y) = \partial_\zeta \textrm{sol}_{K,1}^{(0)}(y)\,.
\eeqa
Next,~let us consider Poisson's equation and its application in electrostatics in various spatial
dimensions~\cite{Ivanenko&Sokolov}.~In case of one spatial dimension,~this yields the electrostatic
potential~which depends linearly on distance from a point-like source.~This~setup is
realised~in~three~spatial dimensions by treating
a uniformly charged plane at distances much smaller than~the~plane's
size.~In case~of~two spatial dimensions, Poisson's equation provides the electrostatic potential
depending logarithmically on distance.~In practice,~this describes a uniformly charged line~at
distances much smaller than its length.~Finally, a point-like electric charge is characterised~by
Coulomb's potential in three spatial dimensions.~In~all of these cases,~the~Green's function~is
functionally given by a Fourier integral with the number of integrations equaling~the
number of spatial dimensions.~The potentials in $3$ and $2$ spatial dimensions
are accordingly~related~as follows:
$1/|\boldsymbol{x}| \propto \partial_{|\boldsymbol{x}|} \ln|\boldsymbol{x}|$.~Therefore,
studying the potential in $2$ spatial dimensions allows~in practice
to determine the potential in $3$ spatial dimensions in theory. 

With this analogy at hand,
one might use empirical results for the two-dimensional~sphere,
or~$\text{CESU}_3$,~to test if $\textrm{sol}_{K,1}^{(0)}(y)$ in $\text{CESU}_3$ gives quantum-particle
dynamics being in agreement with measurements.~If affirmative,~$\textrm{sol}_{K,1}^{(0)}(y)$~in~$\text{CESU}_3$ provides
$\textrm{sol}_{K,1}^{(1)}(y)$ in $\text{CESU}_5$,~in~accord with 
\eqref{eq:alpha1-alpha0}.~Then,~$\textrm{sol}_{K,1}^{(1)}(y)$ in $\text{CESU}_5$~gives
$\textrm{sol}_{K,1}^{(1)}(y)$ in $\text{dS}_4$ by use of the dimensional reduction
and analytic continuation from~Sec.~\ref{sec:metric-tensor}.~Therefore, studying a
Bose-Einstein condensate~on~a 2-sphere might allow to gain some insights into quantum vacuum~in the observable Universe,
as $\text{dS}_4$ approximately models the inflationary and current phases of
cosmic 
evolution~\cite{Mukhanov}.\;\;\;

\section{Concluding remarks}
\label{sec:crs}

Relying on the experimental results involving quantum theory and
gravity, both indirectly and directly, we have put forward the idea that particles' states in curved spacetime should be
\emph{locally} reducible to those based on irreducible unitary representations of the Poincar\'{e} group
\cite{Emelyanov-2020,Emelyanov-2021,Emelyanov-2022a,Emelyanov-2022b}.~The realisation
of this idea naturally involves both~the Einstein equivalence~principle
and the principle of general covariance.~As a result,~theoretical particles~modelled this way
in gravity have properties matching those of observable particles.~This,~as an example,~allows~us
to explain the Colella-Overhauser-Werner observation~\cite{Colella&Overhauser&Werner} as 
being due to gravitational time dilation,~while novel experimental tests~\cite{Emelyanov-2022a,Emelyanov-2022b}
may shed further light on quantum particles in the weak-gravity
regime.

This conceptual idea is mathematically realised through geodesic
distance,~$\sigma(x,X)$.~This is defined through the
classical action for a path between $X$ and $x$,~reaching its extremum~on a geodesic
connecting $X$ and $x$~\cite{DeWitt}.~This is a geometric object being
a zero-rank tensor~with respect to both $x$ and $X$, connected to Riemann normal coordinates
through~\eqref{eq:rnc-gd}~\cite{Ruse}.~This relation reveals the way $\sigma(x,X)$ enters
particles' phase in quantum theory in the Minkowski-spacetime approximation of the observable
Universe -- the approximation used in theoretical particle physics.~Though,~$\exp_K(y)$
is a non-exact solution of the Klein-Gordon equation~in
the Universe owing to non-zero
Riemann tensor.~This entails the replacement~$\exp_K(y) \to \textrm{sol}_K(y)$
in~\eqref{eq:adagger-minkowski} giving~\eqref{eq:adagger-universe}.\,We have succeeded here in
deriving $\textrm{sol}_K(y)$ in AdS, dS, CESU and OESU.
Based
on this result,~we have then derived a \emph{single} solution for
$\textrm{AdS}_{2\alpha+2}$,~$\textrm{dS}_{2\alpha+ 2}$,~$\textrm{CESU}_{2\alpha+3}$
and $\textrm{OESU}_{2\alpha+3}$ with $\alpha \in \{0,\,1\}$.~This~solution,~denoted by
$\textrm{sol}_{K,1}^{(\alpha)}(y)$, is
\begin{enumerate}[label=(\roman*)]
\item tending to positive-frequency plane waves in the limit of vanishing spacetime curvature,
\item invariant under general coordinate transformations,
\item non-perturbative in spacetime curvature,
\item linked to the Wightman functions derived in the literature for the spacetimes studied.
\end{enumerate}
The first property makes $\textrm{sol}_{K,1}^{(\alpha)}(y)$ suitable for the standard applications of quantum~theory 
in particle physics,~while the third property allows then to~gain~insights~into quantum physics
in the strong-gravity regime~\cite{Emelyanov-2020}.

It,~however,~remains to generalise this result to
$\alpha \in \{1/2,\,3/2\}$.~Specifically,~it might be~of interest to jointly treat the pair of an $\text{AdS}_d$ and an $\text{OESU}_{d+1}$,~taking into account
the anti-de Sitter/conformal field theory correspondence~\cite{Maldacena,Witten,Gubser&Klebanov&Polyakov}.~In the case of $\mathcal{N} =4$ supersymmetric Yang-Mills theory in four-dimensional Minkowski spacetime, one would need to
examine the case with $d = 5$ or, equivalently, $\alpha = 3/2$.~Along with this,~it also might~be~of interest to
study a closed $\textrm{ESU}_4$, implying that $\alpha = 1/2$.~This would allow to study quantum-particle~dynamics not only in a (quasi-)static limit of the Oppenheimer-Snyder model of gravitational~collapse
\cite{Oppenheimer&Snyder},
but also in presence of boundary~\cite{Volovik}.

\section*{
ACKNOWLEDGMENTS}

We thank one of the anonymous referees for constructive suggestions~which encouraged~us
to enrich the presentation of our manuscript.~This research has been funded by the Deutsche Forschungsgemeinschaft (DFG,~German Research
Foundation) with the project number 442047500 (SFB 1481).

\end{document}